\documentclass[aps,twocolumn,floats,longbibliography]{revtex4}
\usepackage{graphicx}
\usepackage{dcolumn}
\usepackage{bm}
\usepackage{color}
\usepackage{amsmath}
\usepackage{amssymb}
\usepackage{hyperref}

\begin{document}

\title{One-dimensional Repulsive Fermi Gas in a Tunable Periodic Potential}

\author{$^{1,3}$Sebastiano Pilati}
\author{$^{2}$Luca Barbiero}
\author{$^{5,3,4}$Rosario Fazio}
\author{$^{1}$Luca Dell'Anna}

\affiliation{$^{1}$Dipartimento di Fisica e Astronomia ``Galileo Galilei'' and CNISM, Universit\`a di Padova, I-35131 Padova, Italy}
\affiliation{$^{2}$CNR-IOM DEMOCRITOS Simulation Center, I-34136 Trieste}
\affiliation{$^{3}$Scuola Normale Superiore, I-56126 Pisa, Italy}
\affiliation{$^{4}$Istituto Nanoscienze-CNR, I-56126 Pisa, Italy}
\affiliation{$^{5}$The Abdus Salam International Centre for Theoretical Physics, I-34151 Trieste, Italy}

\begin{abstract}
By using unbiased continuos-space quantum Monte Carlo simulations, we investigate the ground state properties of a one-dimensional repulsive Fermi gas subjected to a commensurate periodic optical lattice (OL) of arbitrary intensity.
The equation of state and the magnetic structure factor are determined as a function of the interaction strength and of the OL intensity. 
In the weak OL limit, Yang's theory for the energy of a homogeneous Fermi gas is recovered. In the opposite limit (deep OL), we analyze the convergence to the Lieb-Wu theory for the Hubbard model, comparing two approaches to map the continuous-space to the discrete-lattice model: the first is based on (noninteracting) Wannier functions, the second effectively takes into account strong-interaction effects within a parabolic approximation of the OL wells.
We find that strong antiferromagnetic correlations emerge in deep OLs, and also in very shallow OLs if the interaction strength approaches the Tonks-Girardeau limit. In deep OLs we find quantitative agreement with density matrix renormalization group calculations for the Hubbard model. The spatial decay of the antiferromagnetic correlations is consistent with quasi long-range order even in shallow OLs, in agreement with previous theories for the half-filled Hubbard model.
\end{abstract}


\maketitle

Making unbiased predictions for the properties of strongly correlated Fermi systems is one of the major challenges in quantum physics research. One dimensional systems play a central role in this context since, on the one hand, correlations effects are more pronounced in low dimensions and, on the other hand, exact results have been derived in a few relevant cases~\cite{giamarchi2004quantum}. Two such cases are the homogeneous Fermi gas, whose exact ground-state energy was first determined by Yang~\cite{yang} via the Bethe Anstatz technique, and the single-band Hubbard model, whose solution was provided by Lieb and Wu~\cite{liebwu}.
These two paradigmatic models describe two opposite limits of realistic physical systems, which in general are neither perfectly homogeneous nor devoid of interband couplings. In the absence of exact analytical theories for the more realistic intermediate regime, developing unbiased computational techniques is of outmost importance.\\
The experiments performed with ultracold atoms trapped in optical lattices (OLs) have emerged as the ideal playground to investigate quantum many-body phenomena in periodic potentials~\cite{bloch2008many}. The intensity of the external periodic field can be easily varied by tuning a laser power, and also the interaction strength can by tuned exploiting Feshbach resonances~\cite{chin}. This has recently allowed the remarkable observation of antiferromagnetic correlations in a controlled experimental setup, both in two and in one dimension~\cite{greif2013short,hart2015observation,parsons2016site,mazurenko2016experimental,brown2016observation,Boll1257,hilker2017revealing}.\\
The bulk of early research activity on OL systems focussed on deep OLs and weak interactions, where single-band tight-binding models are adequate~\cite{jaksch2005cold}. Away from this regime multi-band processes come into play, and the effect of the \emph{independent} tuning of the OL intensity and the interaction strength can be captured only via multi-band or continuous-space models. Recent theoretical and experimental studies have addressed the regime of shallow OLs and strong interactions, investigating intriguing phenomena such as Mott and pinning bosonic localization transitions~\cite{haller2010pinning,pilati2011bosonic,de2012phase,astrakharchik2016one,boeris}, Anderson localization~\cite{boers2007mobility,biddle2009localization,pilati2017localization}, Bose-Glass phases~\cite{PhysRevA.91.043618}, and itinerant ferromagnetism~\cite{dft,pilati2014}.\\

Previous theoretical studies on extended one-dimensional Fermi gases considered either homogeneous continuous-space systems or discrete-lattice models.
In this Rapid Communication, we investigate the ground-state properties of a continuous-space one-dimensional Fermi gas with zero-range repulsive interactions, subjected to a periodic potential (representing an OL) of arbitrary intensity. We focus on a balanced (i.e., unpolarized) two-component mixture at the density of one fermion per well (half filling). The energy and the magnetic structure factor are computed via continuous-space diffusion Monte Carlo (DMC) simulations, which provide unbiased predictions for one-dimensional Fermi systems.\\
We explore the crossover between two opposite limits. For a vanishing OL, we recover the ground state energy of a homogeneous system predicted by Yang; for a deep OL, where the continuous-space system can be mapped to a discrete-lattice model, we inspect the convergence to the Lieb-Wu results for the Hubbard model.
Specifically, we consider two mapping procedures; the first is based on the standard Wannier functions, the second is designed to effectively take into account within an harmonic approximation the higher-orbital effects induced by strong interactions. The regimes where these two mapping procedures become quantitatively accurate are outlined.
Furthermore, the onset of the antiferromagnetic correlations is explored. We find that strong correlations form in deep OLs, where the continuos-space DMC data agree with Hubbard-model results, which we obtain using the density matrix renormalization group (DMRG) method. Interestingly, we find that the correlation amplitude can be large even in very shallow lattices if the interaction strength is tuned close to the infinite repulsive (Tonks-Girardeau) limit.
Both in deep and in shallow OLs the spatial decay of the correlations appears to be consistent with the quasi-long range order predicted by bosonization theories for the half-filled Hubbard model and for the one-dimensional Wigner crystal.\\

%
%
%
%
\begin{figure}
\begin{center}
\includegraphics[width=1.0\columnwidth]{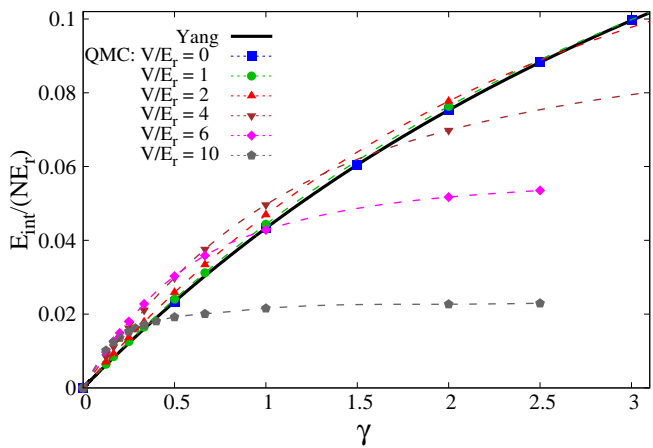}
\caption{(Color online) Ground-state interaction energy per particle $E_{\mathrm{int}}/N=(E-E_{\gamma=0})/N$ as a function of the interaction parameter $\gamma = 2/(n|a_{1D}|)$. $E$ and $E_{\gamma=0}$ are the energies of an interacting and a noninteracting gas in an OL, respectively. The density is fixed at half filling $n=N/(Ld)=1/d$.
Symbols connected by dashed lines correspond to QMC results (system size $L=26$) for different OL intensities $V$, expressed in units of the recoil energy $E_r$. The thick continuous-curve is the Yang's Bethe-Anstatz result~\cite{yang} for the homogeneous Fermi gas ($V=0$).
}
\label{fig1}
\end{center}
\end{figure}

%
We consider a one-dimensional two-component atomic Fermi gas described by the following continuous-space Hamiltonian:
\begin{equation}
\hat{H} = 
       \sum_{i=1             }^{N} \left( -\frac{\hbar^2}{2m}\frac{\mathrm{d}^2}{\mathrm{d}x_{i}^2}  + v(x_i)   \right)
       +  \sum_{i_\uparrow,i_\downarrow}g\delta(x_{i_\uparrow} -x_{i_\downarrow}) 
       \;,
\label{hamiltonian}
\end{equation}
where $\hbar$ is the reduced Planck constant, $m$ is the atomic mass, and the total particle number is $N=N_\uparrow +N_\downarrow$, where $N_\uparrow$ and $N_\downarrow$ are the number of particles of the two components (hereafter referred to as spin-up and spin-down particles).  The index $i=1,\dots,N$ labels all particles (irrespectively of their spin state), while the indices $i_\uparrow$ and $i_\downarrow$  label, respectively, only spin-up and only spin-down particles. We focus on a balanced (unpolarized) mixture of the two components  $N_\uparrow= N_\downarrow=N/2$.
The external potential $v(x)=V \sin^2\left(\pi x/d\right) $ represents the effect of an optical lattice with period $d$ and intensity $V$. The latter will be conveniently expressed in units of the recoil energy $E_r = \hbar^2\pi^2/(2md^2)$. We focus on a half-filled lattice, where the average density is $n = N/(Ld) = 1/d$. The linear system size is $Ld$, being $L$ the number of wells of the OL. This is consistent with the use of periodic boundary conditions.
The interaction strength is fixed by the one-dimensional coupling constant $g=-2\hbar^2/(ma_{1D})$, where $a_{1D}$ is the one-dimensional scattering length. We consider the case of repulsive interactions $g\geqslant 0$. 
In the experiments preformed with atomic clouds confined in tight cigar-shaped waveguides, the coupling constant $g$ can be related to 
the relevant experimental parameters~\cite{olshanii1998atomic}, namely the three-dimensional s-wave scattering length and the radial harmonic confining frequency (assumed to be sufficiently strong to freeze the radial modes). Following the conventional formalism of homogeneous one-dimensional Fermi gases~\cite{astrakharchikgiorgini}, we cast the interaction parameter in the adimensional form $\gamma=2/(n|a_{1D}|)$.\\

%
%
%
%
\begin{figure}
\begin{center}
\includegraphics[width=1.0\columnwidth]{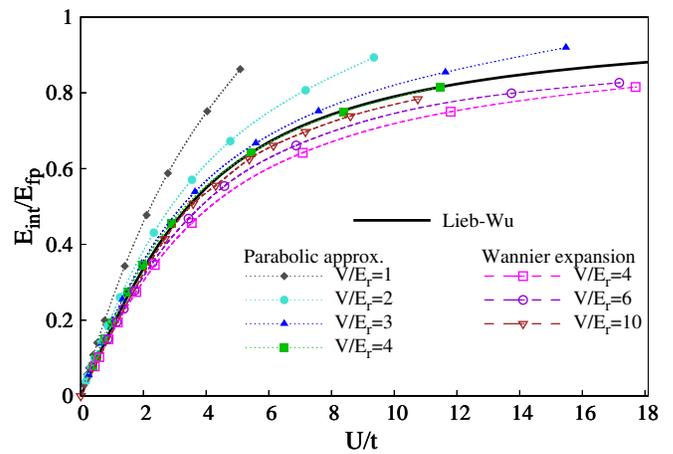}
\caption{(Color online)
Comparison between the ground-state interaction energy of the one-dimensional Hubbard model determined via Bethe Anstatz theory by Lieb and Wu~\cite{liebwu} (continuous thick black curve), and the continuous-space data obtained via DMC simulations. The parameters of the continuous-space model~(\ref{hamiltonian}) are mapped to the Hubbard interaction parameter $U/t$ either using the standard Wannier expansion~\cite{tU} (empty symbols connected by dashed lines) or via a parabolic approximation (solid symbols connected by dotted lines) which effectively accounts for higher-bands effects (see text). Different datasets correspond to different OL intensities $V/E_r$. $E_{\mathrm{fp}}$ is the energy of a fully polarized Fermi gas ($N_{\uparrow}=N$ and $N_{\downarrow}=0$).
}
\label{fig2}
\end{center}
\end{figure}
\begin{figure}
\begin{center}
\includegraphics[width=1.0\columnwidth]{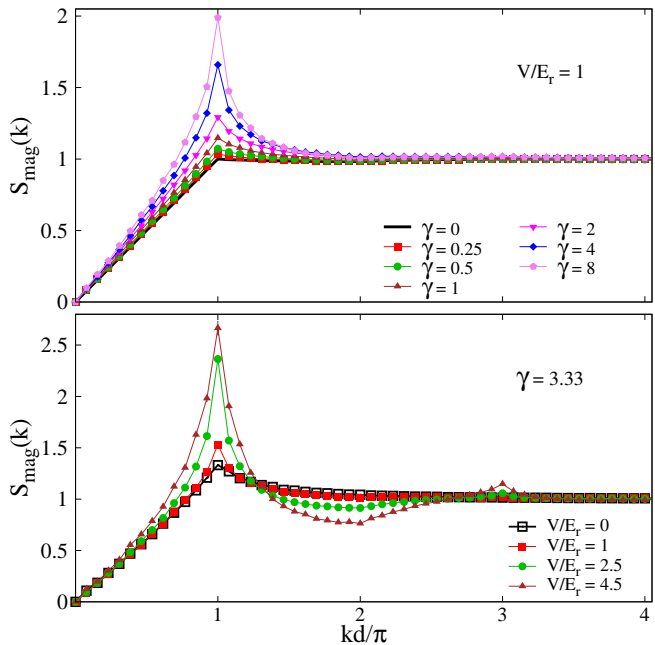}
\caption{(Color online)
Static magnetic structure factor $S_{\mathrm{mag}}(k)$ of the continuous-space model~(\ref{hamiltonian}) as a function of the wave-vector $k$. $d$ is the OL periodicity. The different datasets correspond to different values of the interaction parameter $\gamma$ at the same OL depth $V$ (upper panel), and to different OL intensities $V$ at the same interaction strength $\gamma$ (lower panel). $E_r$ is the recoil energy.  The particle number is $N=26$.
}
\label{fig3}
\end{center}
\end{figure}

 %
\begin{figure}
\begin{center}
\includegraphics[width=1.0\columnwidth]{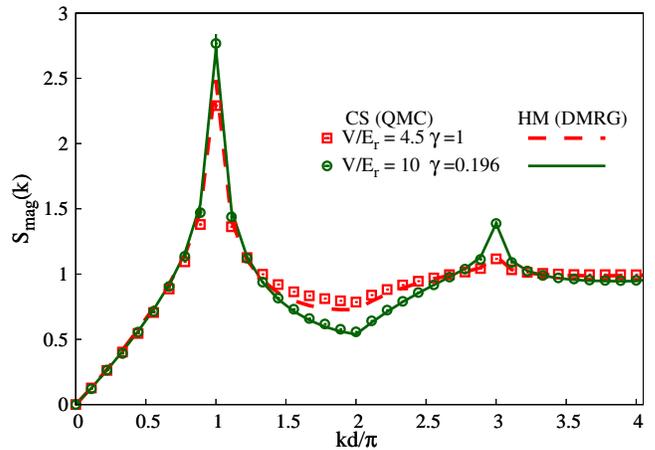}
\caption{(Color online)
Comparison between the static magnetic structure factors $S_{\mathrm{mag}}(k)$ of the Hubbard model (HM), computed via the DMRG method (solid and dashed curves), and of the continuous-space (CS) model, computed via QMC simulations (empty symbols). The (red) squares and the dashed curve correspond to the OL intensity $V/E_r=4.5$ and the continuous-space interaction parameter $\gamma=1$, while the (green) circles and the solid curve to  $V/E_r=10$ and $\gamma =0.196$. These two pairs of continuous-space parameters correspond to the same Hubbard interaction parameter $U/t\cong 4.2$, according to the standard Wannier expansion~\cite{tU}. The Hubbard model results have been converted via eq.~(\ref{skconf}), using the Wannier functions at the corresponding OL intensity.  The particle number is $N=18$.
}
\label{fig4}
\end{center}
\end{figure}
%
\begin{figure}
\begin{center}
\includegraphics[width=1.0\columnwidth]{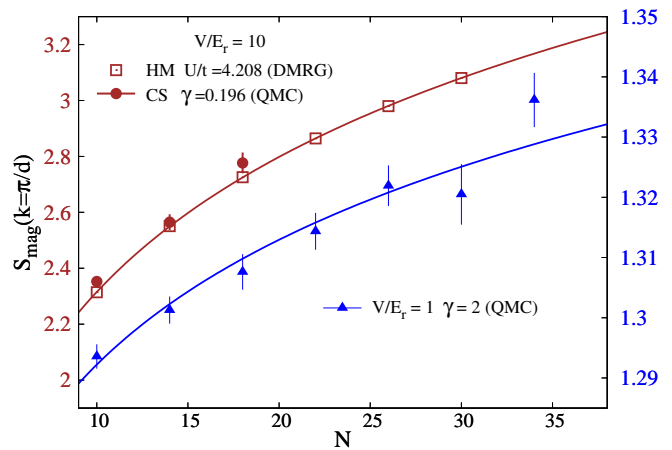}
\caption{(Color online)
Value of the magnetic structure factor $S_{\mathrm{mag}}(k)$ at the peak wave-vector $k=\pi/d$ as a function of the particle number $N$. Upper (brown) datasets correspond to a deep OL of intensity $V/E_r=10$ (referred to the left vertical axis), lower (blue) dataset to a shallow lattice $V/E_r=1$ (referred to the right vertical axis). Full symbols represent QMC data for the continuous-space (CS) model, while empty squares to DMRG data for the Hubbard model (HM), converted via eq.~(\ref{skconf}).
}
\label{fig5}
\end{center}
\end{figure}

%
The ground state properties of the Hamiltonian~(\ref{hamiltonian}) are determined via quantum Monte Carlo (QMC) simulations based on the DMC algorithm~\cite{reynolds1982fixed}. 
While in generic many-fermion systems the sign problem hinders exact QMC simulations, in one dimension this pathology can be circumvented without introducing any systematic approximation since the exact nodal structure is known~\cite{ceperley1991fermion,casula2006ground,casula2008quantum,matveeva2016one}. In order to reduce the statistical fluctuations we employ Jastrow-Slater trial wave functions. The details of our implementation of the DMC algorithm have been reported in Refs.~\cite{pilati2017localization,pilati2014}. 
In order to compute unbiased expectation values of operators that do not commute with the Hamiltonian we employ the standard forward walking technique~\cite{boronat}.\\ 

In Fig.~\ref{fig1} we report the interaction energy per particle $E_{\mathrm{int}}/N=(E-E_{\gamma=0})/N$, where $E$ and $E_{\gamma=0}$ are the total energies of an interacting and of a noninteracting ($\gamma = 0$) gas in an OL, respectively. 
These data correspond to the particle number $N=26$. In fact, by performing a finite-size scaling analysis using particle numbers in the range $18\leqslant N \leqslant 54$, we verified that with $N=26$ the relative error due to the finite system size is below the statistical errorbars in the weak interaction regime, and still below $0.3\%$ in the regime of strong interactions $\gamma\approx 3$.
In the weak OL limit $V\rightarrow 0$, the DMC results converge to the equation of state for a homogenous Fermi gas. This was determined in Ref.~\cite{astrakharchikgiorgini} by numerically solving the set of integral equations obtained by Yang~\cite{yang} via the Bethe Ansatz technique. $E_{\mathrm{int}}$ increases with the interaction strength $\gamma$, but it saturates in the Tongs-Girardeau limit $\gamma \rightarrow \infty$, where the energy of a fully polarized gas is reached. While in a shallow OL ($V \approx E_r$) this saturation occurs only for strong interactions $\gamma \gg 1$, in a deep OL ($V\gg E_r$) it occurs already for intermediate interaction strengths $\gamma \approx 1$, meaning that correlation effects are enhanced in deep OLs compared to shallow OLs.\\

In the deep OL limit, one expects that higher bands become irrelevant if the interaction strength is not strong enough to promote interband transitions.  By expanding the field operator in the basis of (maximally localized) Wannier functions, removing higher-band contributions, and neglecting also beyond nearest-neighbor and interaction-induced processes (e.g., bond-charge interaction), the Hamiltonian~(\ref{hamiltonian}) can be mapped to a discrete single-band lattice model, namely the (one-dimensional) Hubbard model~\cite{jaksch}:
\begin{equation}
\label{HM}
  \hat{H} = -t\sum_{{r,\sigma}}  (\hat{c}_{{r,\sigma}}^{\dagger}\hat{c}^{\phantom{\dagger}}_{{r}+{1},\sigma} + 
  \textrm{h.c}) +  U\sum_{r}{\hat{n}}_{r,\uparrow}{\hat{n}}_{r,\downarrow},
 \end{equation}
\noindent
where $\hat{c}^{{\dagger}}_{r,\sigma}$ ($\hat{c}_{{r,\sigma}}$) creates (destroys) a fermion of spin $\sigma\in{\uparrow,\downarrow}$ at site $r$ (with $r=1, \dots, L$), and ${\hat{n}}_{r,\sigma}=c_{{r,\sigma}}^{\dagger}c_{{r,\sigma}}$ is the corresponding number operator. Consistently with the use of periodic boundary conditions, it is understood that  $\hat{c}^{{\dagger}}_{L+1,\sigma}= \hat{c}^{{\dagger}}_{1,\sigma}$ ($\hat{c}_{{L+1,\sigma}} = \hat{c}_{{1,\sigma}}$).
The hopping energy $t$ and the on-site interaction parameter $U$ can be computed from Wannier functions integrals~\cite{jaksch,astrakharchik2016one} following the standard procedure~\cite{tU}.\\
The zero-temperature equation of state of the Hubbard model~(\ref{HM}) was first determined by Lieb and Wu~\cite{liebwu} using the Bethe Ansatz technique~\cite{LWHalfFilling}.
The comparison displayed in Fig.~\ref{fig2} confirms that the continuous-space data do indeed converge to the Hubbard-model results if the OL is sufficiently deep. At the OL intensity $V/E_r=10$, discrepancies are sizable only for large values of the Hubbard interaction parameter $U/t\gtrsim 10$, which corresponds to the intermediate continuous-space interaction parameter $\gamma \gtrsim 0.5$.\\
Inducing strong-correlation effects in shallower lattices demands larger $\gamma$ values. In this regime interband transitions become relevant; therefore, the mapping to a single-band model based on the standard Wannier function expansion~\cite{tU} is invalid. As an attempt to take orbital excitations into account, we consider a parabolic approximation for the OL wells.  The energy $E_2$ of two interacting opposite-spin fermions in the harmonic well can be exactly computed~\cite{xianlong2006density,E2}.
In the noninteracting case one has $E_2=\hbar\omega$, while in the Tonks-Girardeau ($\gamma\rightarrow \infty$) limit one has $E_2=2\hbar\omega$, as for two spin-aligned fermions. We henceforth define the one-site interaction parameter as the interaction energy $U=E_2-\hbar\omega$.
In correspondence with the parabolic approximation for the interaction energy, we compute the hopping energy $t$ using the well-known approximation $-$ valid in the deep OL limit $V/E_r\gg1$ $-$ for (one fourth of) the bandwidth of the lowest band in the one-dimensional Mathieu equation, namely $t=4\pi^{-1/2} E_r (V/E_r)^{3/4}\exp\left(-2\sqrt{V/E_r}\right)$~\cite{bloch2008many}. This formula accounts to leading-order for the splitting of the harmonic-oscillator energy-levels due to tunneling~\cite{ma14ller2006introduction,krutitsky2016ultracold}.
The comparison of Fig.~\ref{fig2} shows that with this effective mapping procedure agreement between continuos-space and Hubbard-model data is obtained already at the moderate lattice depth $V/E_r = 4$, even when the continuous-space interaction parameter is as large as $\gamma =5$ (where $U/t \simeq11.45$, according to this second mapping criterion).\\

Beyond the equation of state, we investigate how the antiferromagnetic correlations depend of the OL intensity and on the interactions strength. To quantify these correlations, we compute via DMC simulations the static magnetic structure factor of the continuous-space model: $S_{\mathrm{mag}}(k) = \left< \rho_{\mathrm{mag}}(k) \rho_{\mathrm{mag}}(-k) \right>$, where
 $\rho_{\mathrm{mag}}(k)$ is the Fourier transform of the spin density operator.
%
The results for $S_{\mathrm{mag}}(k)$ are shown in Fig.~\ref{fig3}. In the upper panel, the different datasets correspond to different values of the interaction strength at the same OL intensity; in the lower panel, the OL intensity varies while the interaction strength is fixed. The peak of $S_{\mathrm{mag}}(k)$ at $k=\pi/d$ signals antiferromagnetic correlations commensurate with the OL. One notices that such correlations can be amplified both by making the OL deeper and by increasing the interaction strength. In a deep OL of intensity $V/E_r=10$, where the single-band description is applicable (see below), strong correlations emerge already at the moderate interaction strength $\gamma\simeq 0.2$ (see Fig.~\ref{fig4}). However, even in OLs as shallow as $V/E_r=1$, strong correlations form if the interaction parameter is close to the Tonks-Girardeau regime $\gamma \gg 1$, meaning that antiferromagnetism occurs also well beyond the tight-binding regime.\\
 In order to analyze the convergence to the single-band limit, we make comparison with DMRG~\cite{white1992density} results for the Hubbard model. Specifically, we compute via the DMRG method the spin-spin correlation function $g(r_1,r_2)= \left<\hat{S}^z_{r_1}\hat{S}^z_{r_2}\right>$, where the spin density operator is $\hat{S}^z_{r}={\hat{n}}_{r,\uparrow}-{\hat{n}}_{r,\downarrow}$~\cite{DMRG}. The Hubbard model results can be compared with the continuous-space magnetic structure factor using the following transformation (valid in the tight-binding limit)~\cite{krutitsky2016ultracold}:
 \begin{equation}
 \label{skconf}
 S_{\mathrm{mag}}(k) = 1+G^2(k)\left[ \tilde{S}_{\mathrm{mag}}(k) -1 \right],
 \end{equation}
 where $\tilde{S}_{\mathrm{mag}}(k)= 1/N\sum_{r_1,r_2} g(r_1,r_2)\exp\left[ i k (r_1-r_2)\right]$, and $G(k) = \int |w_r(x)|^2 \exp(-ikx) \mathrm{d} x$. It is worth stressing that in one dimension the spin density $\left<\hat{S}^z_{r}\right>$ is strictly zero as a consequence of the Mermin-Wagner theorem.
The comparison between continuous-space and Hubbard model data is displayed in Fig.~\ref{fig4}. We find that at the moderate OL depth $V/E_r=4.5$ sizable discrepancies still persist, but in a deeper OL of intensity $V/E_r=10$ precise matching is achieved.\\
While the antiferromagnetic correlations are pronounced at large $V/E_r$ and/or strong $\gamma$, the Mermin-Wagner theorem implies the absence of proper antiferromagnetic long-range order. It is therefore interesting to inspect how the spin-spin correlations decay at long distance. This problem has been the subject of thorough theoretical investigations in the context of the Hubbard model~\cite{schulz1990correlation,frahm1990critical,parola1990asymptotic,imada1992crossover,sandvik1993quantum,qin1995density,iino1995crossover}, and also in the context of continuous-space systems with (long-range) Coulomb repulsion (and no external potentials), which at low density form a Wigner crystal characterized by quasi long-range density-density correlations~\cite{schulz1993wigner}. While at short distance the spin-spin correlation functions of the two systems are not identical, their long-distance behavior is fixed to leading order by the same power-law $g(r_1,r_2)\sim \left|r_1-r_2\right|^{-(1+\alpha)}$ ($g(x_1,x_2)\sim \left|x_1-x_2\right|^{-(1+\alpha)}$ in the continuous-space notation), with logarithmic corrections~\cite{HMcorr}. The value of the exponent $\alpha=0$ (at finite $U$), which was determined via the bosonization technique~\cite{schulz1990correlation,frahm1990critical,schulz1993wigner}, indicates quasi long-range spin order. This behavior implies, to leading order, a logarithmic divergence of the peak value $S_{\mathrm{mag}}(k=\pi/d)$ with the particle number $N$. It is worth mentioning that away from half filling the spin-spin correlations of the  Hubbard model have a short-range character fixed by the exponent $\alpha=1/2$, implying that the peak value $S_{\mathrm{mag}}(k=\pi/d)$ is finite in the thermodynamic limit.
The system-size dependence of both the DMC continuous-space and the DMRG Hubbard model data $S_{\mathrm{mag}}(k=\pi/d)$ is displayed in Fig.~\ref{fig5}. The Hubbard model data (which agree with continuous-space DMC data at $V/E_r=10$) are well reproduced by a logarithmic fitting function of the type $f(N) = a + b\ln(N)$ ($a$ and $b$ are the fitting parameters), in agreement with the bosonization theory. Also in the shallow lattice $V/E_r=1$ we observe a slow increase of $S_{\mathrm{mag}}(k=\pi/d)$ with system size, again well described by the logarithmic fit $f(N)$. While smaller statistical error-bars and larger system sizes would be needed to rule out different functional forms, the available data suggest that even in a very shallow OL, where single-band models are inadequate, the asymptotic decay of the spin-spin correlation function displays quasi long-range order.\\

In conclusion, we determined via unbiased continuous-space DMC simulations the zero-temperature equation of state and the static magnetic structure factor of a one-dimensional Fermi gas in a half-filled OL of arbitrary intensity. 
We analysed the accuracy of two procedures to map the continuous-space Hamiltonian to the Hubbard model. The first $-$ which turns out to be very accurate if $V/E_r\gtrsim 10$ and $\gamma\lesssim 0.5$ $-$ is based on the standard Wannier function expansion. The second $-$ which is accurate even in the presence of interactions as strong as $\gamma \approx 5$ $-$ is based on a parabolic approximation of the OL wells. This analysis quantifies to what extent OL experiments can be described via single-band lattice models.\\
We shed light on how antiferromagnetic correlations emerge both in deep and in shallow OLs. In the former case, the correlations have a quasi long-range character, in agreement with Hubbard model predictions (bosonization and DMRG calculations), and they are pronounced already in the moderate interaction regime $\gamma \approx 0.2$. Remarkably, also in the latter case strong correlations occur (again, consistent with quasi long-range order) for interaction strengths $\gamma \gg 1$, where multiband effects are important.\\
While previous theoretical studies on confined one-dimensional fermions addressed the case of (typically small) harmonic traps~\cite{PhysRevA.88.033607,volosniev2014strongly,Matveeva,Lindgren,PhysRevA.87.060502,PhysRevLett.111.045302,PhysRevA.91.043634,PhysRevA.89.023603,loft2015variational,grining2015crossover,twoflavour,PhysRevA.91.053618,dehkharghani2016impenetrable,PhysRevA.91.013620,Levinsene1500197}, in this Rapid Communication we considered a commensurate periodic potential, using sufficiently large system sizes to inspect the thermodynamic limit. On the one hand, this study provides new unbiased predictions for a paradigmatic model of strongly-correlated Fermi systems, which interpolates between Yang theory of the homogeneous Fermi gas and Lieb-Wu theory of the Hubbard model; on the other hand, it serves as a guide for possible new cold-atoms experiments aiming at observing antiferromagnetism beyond the tight-binding regime~\cite{casula2006ground,Stella}.\\

We acknowledge interesting discussions with F. Ancilotto.
S. P. acknowledges financial support from the BIRD 2016 project ``Superfluidit\`a in gas fermionici
ultrafreddi in due dimensioni'' of the University of Padova, and the EU-H2020 project No. 641122 QUIC - Quantum simulations of insulators and conductors.
L. B. acknowledges support from the European STREP MatterWave, Karma cluster in Trieste for CPU time and  University of Padova for kind hospitality.

\begin{thebibliography}{66}%
\makeatletter
\providecommand \@ifxundefined [1]{%
 \@ifx{#1\undefined}
}%
\providecommand \@ifnum [1]{%
 \ifnum #1\expandafter \@firstoftwo
 \else \expandafter \@secondoftwo
 \fi
}%
\providecommand \@ifx [1]{%
 \ifx #1\expandafter \@firstoftwo
 \else \expandafter \@secondoftwo
 \fi
}%
\providecommand \natexlab [1]{#1}%
\providecommand \enquote  [1]{``#1''}%
\providecommand \bibnamefont  [1]{#1}%
\providecommand \bibfnamefont [1]{#1}%
\providecommand \citenamefont [1]{#1}%
\providecommand \href@noop [0]{\@secondoftwo}%
\providecommand \href [0]{\begingroup \@sanitize@url \@href}%
\providecommand \@href[1]{\@@startlink{#1}\@@href}%
\providecommand \@@href[1]{\endgroup#1\@@endlink}%
\providecommand \@sanitize@url [0]{\catcode `\\12\catcode `\$12\catcode
  `\&12\catcode `\#12\catcode `\^12\catcode `\_12\catcode `\%12\relax}%
\providecommand \@@startlink[1]{}%
\providecommand \@@endlink[0]{}%
\providecommand \url  [0]{\begingroup\@sanitize@url \@url }%
\providecommand \@url [1]{\endgroup\@href {#1}{\urlprefix }}%
\providecommand \urlprefix  [0]{URL }%
\providecommand \Eprint [0]{\href }%
\providecommand \doibase [0]{http://dx.doi.org/}%
\providecommand \selectlanguage [0]{\@gobble}%
\providecommand \bibinfo  [0]{\@secondoftwo}%
\providecommand \bibfield  [0]{\@secondoftwo}%
\providecommand \translation [1]{[#1]}%
\providecommand \BibitemOpen [0]{}%
\providecommand \bibitemStop [0]{}%
\providecommand \bibitemNoStop [0]{.\EOS\space}%
\providecommand \EOS [0]{\spacefactor3000\relax}%
\providecommand \BibitemShut  [1]{\csname bibitem#1\endcsname}%
\let\auto@bib@innerbib\@empty
\bibitem [{\citenamefont {Giamarchi}(2004)}]{giamarchi2004quantum}%
  \BibitemOpen
  \bibfield  {author} {\bibinfo {author} {\bibfnamefont {T.}~\bibnamefont
  {Giamarchi}},\ }\href@noop {} {\emph {\bibinfo {title} {Quantum physics in
  one dimension}}}\ (\bibinfo  {publisher} {Oxford university press},\ \bibinfo
  {year} {2004})\BibitemShut {NoStop}%
\bibitem [{\citenamefont {Yang}(1967)}]{yang}%
  \BibitemOpen
  \bibfield  {author} {\bibinfo {author} {\bibfnamefont {C.-N.}\ \bibnamefont
  {Yang}},\ }\href@noop {} {\bibfield  {journal} {\bibinfo  {journal} {Phys.
  Rev. Lett.}\ }\textbf {\bibinfo {volume} {19}},\ \bibinfo {pages} {1312}
  (\bibinfo {year} {1967})}\BibitemShut {NoStop}%
\bibitem [{\citenamefont {Lieb}\ and\ \citenamefont {Wu}(1968)}]{liebwu}%
  \BibitemOpen
  \bibfield  {author} {\bibinfo {author} {\bibfnamefont {E.~H.}\ \bibnamefont
  {Lieb}}\ and\ \bibinfo {author} {\bibfnamefont {F.~Y.}\ \bibnamefont {Wu}},\
  }\href@noop {} {\bibfield  {journal} {\bibinfo  {journal} {Phys. Rev. Lett.}\
  }\textbf {\bibinfo {volume} {20}},\ \bibinfo {pages} {1445} (\bibinfo {year}
  {1968})}\BibitemShut {NoStop}%
\bibitem [{\citenamefont {Bloch}\ \emph {et~al.}(2008)\citenamefont {Bloch},
  \citenamefont {Dalibard},\ and\ \citenamefont {Zwerger}}]{bloch2008many}%
  \BibitemOpen
  \bibfield  {author} {\bibinfo {author} {\bibfnamefont {I.}~\bibnamefont
  {Bloch}}, \bibinfo {author} {\bibfnamefont {J.}~\bibnamefont {Dalibard}}, \
  and\ \bibinfo {author} {\bibfnamefont {W.}~\bibnamefont {Zwerger}},\
  }\href@noop {} {\bibfield  {journal} {\bibinfo  {journal} {Rev. Mod. Phys.}\
  }\textbf {\bibinfo {volume} {80}},\ \bibinfo {pages} {885} (\bibinfo {year}
  {2008})}\BibitemShut {NoStop}%
\bibitem [{\citenamefont {Chin}\ \emph {et~al.}(2010)\citenamefont {Chin},
  \citenamefont {Grimm}, \citenamefont {Julienne},\ and\ \citenamefont
  {Tiesinga}}]{chin}%
  \BibitemOpen
  \bibfield  {author} {\bibinfo {author} {\bibfnamefont {C.}~\bibnamefont
  {Chin}}, \bibinfo {author} {\bibfnamefont {R.}~\bibnamefont {Grimm}},
  \bibinfo {author} {\bibfnamefont {P.}~\bibnamefont {Julienne}}, \ and\
  \bibinfo {author} {\bibfnamefont {E.}~\bibnamefont {Tiesinga}},\ }\href@noop
  {} {\bibfield  {journal} {\bibinfo  {journal} {Rev. Mod. Phys.}\ }\textbf
  {\bibinfo {volume} {82}},\ \bibinfo {pages} {1225} (\bibinfo {year}
  {2010})}\BibitemShut {NoStop}%
\bibitem [{\citenamefont {Greif}\ \emph {et~al.}(2013)\citenamefont {Greif},
  \citenamefont {Uehlinger}, \citenamefont {Jotzu}, \citenamefont {Tarruell},\
  and\ \citenamefont {Esslinger}}]{greif2013short}%
  \BibitemOpen
  \bibfield  {author} {\bibinfo {author} {\bibfnamefont {D.}~\bibnamefont
  {Greif}}, \bibinfo {author} {\bibfnamefont {T.}~\bibnamefont {Uehlinger}},
  \bibinfo {author} {\bibfnamefont {G.}~\bibnamefont {Jotzu}}, \bibinfo
  {author} {\bibfnamefont {L.}~\bibnamefont {Tarruell}}, \ and\ \bibinfo
  {author} {\bibfnamefont {T.}~\bibnamefont {Esslinger}},\ }\href@noop {}
  {\bibfield  {journal} {\bibinfo  {journal} {Science}\ }\textbf {\bibinfo
  {volume} {340}},\ \bibinfo {pages} {1307} (\bibinfo {year}
  {2013})}\BibitemShut {NoStop}%
\bibitem [{\citenamefont {Hart}\ \emph {et~al.}(2015)\citenamefont {Hart},
  \citenamefont {Duarte}, \citenamefont {Yang}, \citenamefont {Liu},
  \citenamefont {Paiva}, \citenamefont {Khatami}, \citenamefont {Scalettar},
  \citenamefont {Trivedi}, \citenamefont {Huse},\ and\ \citenamefont
  {Hulet}}]{hart2015observation}%
  \BibitemOpen
  \bibfield  {author} {\bibinfo {author} {\bibfnamefont {R.~A.}\ \bibnamefont
  {Hart}}, \bibinfo {author} {\bibfnamefont {P.~M.}\ \bibnamefont {Duarte}},
  \bibinfo {author} {\bibfnamefont {T.-L.}\ \bibnamefont {Yang}}, \bibinfo
  {author} {\bibfnamefont {X.}~\bibnamefont {Liu}}, \bibinfo {author}
  {\bibfnamefont {T.}~\bibnamefont {Paiva}}, \bibinfo {author} {\bibfnamefont
  {E.}~\bibnamefont {Khatami}}, \bibinfo {author} {\bibfnamefont {R.~T.}\
  \bibnamefont {Scalettar}}, \bibinfo {author} {\bibfnamefont {N.}~\bibnamefont
  {Trivedi}}, \bibinfo {author} {\bibfnamefont {D.~A.}\ \bibnamefont {Huse}}, \
  and\ \bibinfo {author} {\bibfnamefont {R.~G.}\ \bibnamefont {Hulet}},\
  }\href@noop {} {\bibfield  {journal} {\bibinfo  {journal} {Nature}\ }\textbf
  {\bibinfo {volume} {519}},\ \bibinfo {pages} {211} (\bibinfo {year}
  {2015})}\BibitemShut {NoStop}%
\bibitem [{\citenamefont {Parsons}\ \emph {et~al.}(2016)\citenamefont
  {Parsons}, \citenamefont {Mazurenko}, \citenamefont {Chiu}, \citenamefont
  {Ji}, \citenamefont {Greif},\ and\ \citenamefont
  {Greiner}}]{parsons2016site}%
  \BibitemOpen
  \bibfield  {author} {\bibinfo {author} {\bibfnamefont {M.~F.}\ \bibnamefont
  {Parsons}}, \bibinfo {author} {\bibfnamefont {A.}~\bibnamefont {Mazurenko}},
  \bibinfo {author} {\bibfnamefont {C.~S.}\ \bibnamefont {Chiu}}, \bibinfo
  {author} {\bibfnamefont {G.}~\bibnamefont {Ji}}, \bibinfo {author}
  {\bibfnamefont {D.}~\bibnamefont {Greif}}, \ and\ \bibinfo {author}
  {\bibfnamefont {M.}~\bibnamefont {Greiner}},\ }\href@noop {} {\bibfield
  {journal} {\bibinfo  {journal} {Science}\ }\textbf {\bibinfo {volume}
  {353}},\ \bibinfo {pages} {1253} (\bibinfo {year} {2016})}\BibitemShut
  {NoStop}%
\bibitem [{\citenamefont {Mazurenko}\ \emph {et~al.}(2016)\citenamefont
  {Mazurenko}, \citenamefont {Chiu}, \citenamefont {Ji}, \citenamefont
  {Parsons}, \citenamefont {Kan{\'a}sz-Nagy}, \citenamefont {Schmidt},
  \citenamefont {Grusdt}, \citenamefont {Demler}, \citenamefont {Greif},\ and\
  \citenamefont {Greiner}}]{mazurenko2016experimental}%
  \BibitemOpen
  \bibfield  {author} {\bibinfo {author} {\bibfnamefont {A.}~\bibnamefont
  {Mazurenko}}, \bibinfo {author} {\bibfnamefont {C.~S.}\ \bibnamefont {Chiu}},
  \bibinfo {author} {\bibfnamefont {G.}~\bibnamefont {Ji}}, \bibinfo {author}
  {\bibfnamefont {M.~F.}\ \bibnamefont {Parsons}}, \bibinfo {author}
  {\bibfnamefont {M.}~\bibnamefont {Kan{\'a}sz-Nagy}}, \bibinfo {author}
  {\bibfnamefont {R.}~\bibnamefont {Schmidt}}, \bibinfo {author} {\bibfnamefont
  {F.}~\bibnamefont {Grusdt}}, \bibinfo {author} {\bibfnamefont
  {E.}~\bibnamefont {Demler}}, \bibinfo {author} {\bibfnamefont
  {D.}~\bibnamefont {Greif}}, \ and\ \bibinfo {author} {\bibfnamefont
  {M.}~\bibnamefont {Greiner}},\ }\href@noop {} {\bibfield  {journal} {\bibinfo
   {journal} {arXiv:1612.08436}\ } (\bibinfo {year} {2016})}\BibitemShut
  {NoStop}%
\bibitem [{\citenamefont {Brown}\ \emph {et~al.}(2016)\citenamefont {Brown},
  \citenamefont {Mitra}, \citenamefont {Guardado-Sanchez}, \citenamefont
  {Schau{\ss}}, \citenamefont {Kondov}, \citenamefont {Khatami}, \citenamefont
  {Paiva}, \citenamefont {Trivedi}, \citenamefont {Huse},\ and\ \citenamefont
  {Bakr}}]{brown2016observation}%
  \BibitemOpen
  \bibfield  {author} {\bibinfo {author} {\bibfnamefont {P.~T.}\ \bibnamefont
  {Brown}}, \bibinfo {author} {\bibfnamefont {D.}~\bibnamefont {Mitra}},
  \bibinfo {author} {\bibfnamefont {E.}~\bibnamefont {Guardado-Sanchez}},
  \bibinfo {author} {\bibfnamefont {P.}~\bibnamefont {Schau{\ss}}}, \bibinfo
  {author} {\bibfnamefont {S.~S.}\ \bibnamefont {Kondov}}, \bibinfo {author}
  {\bibfnamefont {E.}~\bibnamefont {Khatami}}, \bibinfo {author} {\bibfnamefont
  {T.}~\bibnamefont {Paiva}}, \bibinfo {author} {\bibfnamefont
  {N.}~\bibnamefont {Trivedi}}, \bibinfo {author} {\bibfnamefont {D.~A.}\
  \bibnamefont {Huse}}, \ and\ \bibinfo {author} {\bibfnamefont {W.~S.}\
  \bibnamefont {Bakr}},\ }\href@noop {} {\bibfield  {journal} {\bibinfo
  {journal} {arXiv:1612.07746}\ } (\bibinfo {year} {2016})}\BibitemShut
  {NoStop}%
\bibitem [{\citenamefont {Boll}\ \emph {et~al.}(2016)\citenamefont {Boll},
  \citenamefont {Hilker}, \citenamefont {Salomon}, \citenamefont {Omran},
  \citenamefont {Nespolo}, \citenamefont {Pollet}, \citenamefont {Bloch},\ and\
  \citenamefont {Gross}}]{Boll1257}%
  \BibitemOpen
  \bibfield  {author} {\bibinfo {author} {\bibfnamefont {M.}~\bibnamefont
  {Boll}}, \bibinfo {author} {\bibfnamefont {T.~A.}\ \bibnamefont {Hilker}},
  \bibinfo {author} {\bibfnamefont {G.}~\bibnamefont {Salomon}}, \bibinfo
  {author} {\bibfnamefont {A.}~\bibnamefont {Omran}}, \bibinfo {author}
  {\bibfnamefont {J.}~\bibnamefont {Nespolo}}, \bibinfo {author} {\bibfnamefont
  {L.}~\bibnamefont {Pollet}}, \bibinfo {author} {\bibfnamefont
  {I.}~\bibnamefont {Bloch}}, \ and\ \bibinfo {author} {\bibfnamefont
  {C.}~\bibnamefont {Gross}},\ }\href@noop {} {\bibfield  {journal} {\bibinfo
  {journal} {Science}\ }\textbf {\bibinfo {volume} {353}},\ \bibinfo {pages}
  {1257} (\bibinfo {year} {2016})}\BibitemShut {NoStop}%
\bibitem [{\citenamefont {Hilker}\ \emph {et~al.}(2017)\citenamefont {Hilker},
  \citenamefont {Salomon}, \citenamefont {Grusdt}, \citenamefont {Omran},
  \citenamefont {Boll}, \citenamefont {Demler}, \citenamefont {Bloch},\ and\
  \citenamefont {Gross}}]{hilker2017revealing}%
  \BibitemOpen
  \bibfield  {author} {\bibinfo {author} {\bibfnamefont {T.~A.}\ \bibnamefont
  {Hilker}}, \bibinfo {author} {\bibfnamefont {G.}~\bibnamefont {Salomon}},
  \bibinfo {author} {\bibfnamefont {F.}~\bibnamefont {Grusdt}}, \bibinfo
  {author} {\bibfnamefont {A.}~\bibnamefont {Omran}}, \bibinfo {author}
  {\bibfnamefont {M.}~\bibnamefont {Boll}}, \bibinfo {author} {\bibfnamefont
  {E.}~\bibnamefont {Demler}}, \bibinfo {author} {\bibfnamefont
  {I.}~\bibnamefont {Bloch}}, \ and\ \bibinfo {author} {\bibfnamefont
  {C.}~\bibnamefont {Gross}},\ }\href@noop {} {\bibfield  {journal} {\bibinfo
  {journal} {arXiv:1702.00642}\ } (\bibinfo {year} {2017})}\BibitemShut
  {NoStop}%
\bibitem [{\citenamefont {Jaksch}\ and\ \citenamefont
  {Zoller}(2005)}]{jaksch2005cold}%
  \BibitemOpen
  \bibfield  {author} {\bibinfo {author} {\bibfnamefont {D.}~\bibnamefont
  {Jaksch}}\ and\ \bibinfo {author} {\bibfnamefont {P.}~\bibnamefont
  {Zoller}},\ }\href@noop {} {\bibfield  {journal} {\bibinfo  {journal} {Ann.
  Phys. (N.Y.)}\ }\textbf {\bibinfo {volume} {315}},\ \bibinfo {pages} {52}
  (\bibinfo {year} {2005})}\BibitemShut {NoStop}%
\bibitem [{\citenamefont {Haller}\ \emph {et~al.}(2010)\citenamefont {Haller},
  \citenamefont {Hart}, \citenamefont {Mark}, \citenamefont {Danzl},
  \citenamefont {Reichs{\"o}llner}, \citenamefont {Gustavsson}, \citenamefont
  {Dalmonte}, \citenamefont {Pupillo},\ and\ \citenamefont
  {N{\"a}gerl}}]{haller2010pinning}%
  \BibitemOpen
  \bibfield  {author} {\bibinfo {author} {\bibfnamefont {E.}~\bibnamefont
  {Haller}}, \bibinfo {author} {\bibfnamefont {R.}~\bibnamefont {Hart}},
  \bibinfo {author} {\bibfnamefont {M.~J.}\ \bibnamefont {Mark}}, \bibinfo
  {author} {\bibfnamefont {J.~G.}\ \bibnamefont {Danzl}}, \bibinfo {author}
  {\bibfnamefont {L.}~\bibnamefont {Reichs{\"o}llner}}, \bibinfo {author}
  {\bibfnamefont {M.}~\bibnamefont {Gustavsson}}, \bibinfo {author}
  {\bibfnamefont {M.}~\bibnamefont {Dalmonte}}, \bibinfo {author}
  {\bibfnamefont {G.}~\bibnamefont {Pupillo}}, \ and\ \bibinfo {author}
  {\bibfnamefont {H.-C.}\ \bibnamefont {N{\"a}gerl}},\ }\href@noop {}
  {\bibfield  {journal} {\bibinfo  {journal} {Nature}\ }\textbf {\bibinfo
  {volume} {466}},\ \bibinfo {pages} {597} (\bibinfo {year}
  {2010})}\BibitemShut {NoStop}%
\bibitem [{\citenamefont {Pilati}\ and\ \citenamefont
  {Troyer}(2011)}]{pilati2011bosonic}%
  \BibitemOpen
  \bibfield  {author} {\bibinfo {author} {\bibfnamefont {S.}~\bibnamefont
  {Pilati}}\ and\ \bibinfo {author} {\bibfnamefont {M.}~\bibnamefont
  {Troyer}},\ }\href@noop {} {\bibfield  {journal} {\bibinfo  {journal} {Phys.
  Rev. Lett.}\ }\textbf {\bibinfo {volume} {108}},\ \bibinfo {pages} {155301}
  (\bibinfo {year} {2011})}\BibitemShut {NoStop}%
\bibitem [{\citenamefont {De~Soto}\ and\ \citenamefont
  {Gordillo}(2012)}]{de2012phase}%
  \BibitemOpen
  \bibfield  {author} {\bibinfo {author} {\bibfnamefont {F.}~\bibnamefont
  {De~Soto}}\ and\ \bibinfo {author} {\bibfnamefont {M.}~\bibnamefont
  {Gordillo}},\ }\href@noop {} {\bibfield  {journal} {\bibinfo  {journal}
  {Phys. Rev. A}\ }\textbf {\bibinfo {volume} {85}},\ \bibinfo {pages} {013607}
  (\bibinfo {year} {2012})}\BibitemShut {NoStop}%
\bibitem [{\citenamefont {Astrakharchik}\ \emph {et~al.}(2016)\citenamefont
  {Astrakharchik}, \citenamefont {Krutitsky}, \citenamefont {Lewenstein},\ and\
  \citenamefont {Mazzanti}}]{astrakharchik2016one}%
  \BibitemOpen
  \bibfield  {author} {\bibinfo {author} {\bibfnamefont {G.~E.}\ \bibnamefont
  {Astrakharchik}}, \bibinfo {author} {\bibfnamefont {K.~V.}\ \bibnamefont
  {Krutitsky}}, \bibinfo {author} {\bibfnamefont {M.}~\bibnamefont
  {Lewenstein}}, \ and\ \bibinfo {author} {\bibfnamefont {F.}~\bibnamefont
  {Mazzanti}},\ }\href@noop {} {\bibfield  {journal} {\bibinfo  {journal}
  {Phys. Rev. A}\ }\textbf {\bibinfo {volume} {93}},\ \bibinfo {pages} {021605}
  (\bibinfo {year} {2016})}\BibitemShut {NoStop}%
\bibitem [{\citenamefont {Bo{\'e}ris}\ \emph {et~al.}(2016)\citenamefont
  {Bo{\'e}ris}, \citenamefont {Gori}, \citenamefont {Hoogerland}, \citenamefont
  {Kumar}, \citenamefont {Lucioni}, \citenamefont {Tanzi}, \citenamefont
  {Inguscio}, \citenamefont {Giamarchi}, \citenamefont {D'Errico},
  \citenamefont {Carleo} \emph {et~al.}}]{boeris}%
  \BibitemOpen
  \bibfield  {author} {\bibinfo {author} {\bibfnamefont {G.}~\bibnamefont
  {Bo{\'e}ris}}, \bibinfo {author} {\bibfnamefont {L.}~\bibnamefont {Gori}},
  \bibinfo {author} {\bibfnamefont {M.~D.}\ \bibnamefont {Hoogerland}},
  \bibinfo {author} {\bibfnamefont {A.}~\bibnamefont {Kumar}}, \bibinfo
  {author} {\bibfnamefont {E.}~\bibnamefont {Lucioni}}, \bibinfo {author}
  {\bibfnamefont {L.}~\bibnamefont {Tanzi}}, \bibinfo {author} {\bibfnamefont
  {M.}~\bibnamefont {Inguscio}}, \bibinfo {author} {\bibfnamefont
  {T.}~\bibnamefont {Giamarchi}}, \bibinfo {author} {\bibfnamefont
  {C.}~\bibnamefont {D'Errico}}, \bibinfo {author} {\bibfnamefont
  {G.}~\bibnamefont {Carleo}},  \emph {et~al.},\ }\href@noop {} {\bibfield
  {journal} {\bibinfo  {journal} {Phys. Rev. A}\ }\textbf {\bibinfo {volume}
  {93}},\ \bibinfo {pages} {011601} (\bibinfo {year} {2016})}\BibitemShut
  {NoStop}%
\bibitem [{\citenamefont {Boers}\ \emph {et~al.}(2007)\citenamefont {Boers},
  \citenamefont {Goedeke}, \citenamefont {Hinrichs},\ and\ \citenamefont
  {Holthaus}}]{boers2007mobility}%
  \BibitemOpen
  \bibfield  {author} {\bibinfo {author} {\bibfnamefont {D.~J.}\ \bibnamefont
  {Boers}}, \bibinfo {author} {\bibfnamefont {B.}~\bibnamefont {Goedeke}},
  \bibinfo {author} {\bibfnamefont {D.}~\bibnamefont {Hinrichs}}, \ and\
  \bibinfo {author} {\bibfnamefont {M.}~\bibnamefont {Holthaus}},\ }\href@noop
  {} {\bibfield  {journal} {\bibinfo  {journal} {Phys. Rev. A}\ }\textbf
  {\bibinfo {volume} {75}},\ \bibinfo {pages} {063404} (\bibinfo {year}
  {2007})}\BibitemShut {NoStop}%
\bibitem [{\citenamefont {Biddle}\ \emph {et~al.}(2009)\citenamefont {Biddle},
  \citenamefont {Wang}, \citenamefont {Priour~Jr},\ and\ \citenamefont
  {Sarma}}]{biddle2009localization}%
  \BibitemOpen
  \bibfield  {author} {\bibinfo {author} {\bibfnamefont {J.}~\bibnamefont
  {Biddle}}, \bibinfo {author} {\bibfnamefont {B.}~\bibnamefont {Wang}},
  \bibinfo {author} {\bibfnamefont {D.}~\bibnamefont {Priour~Jr}}, \ and\
  \bibinfo {author} {\bibfnamefont {S.~D.}\ \bibnamefont {Sarma}},\ }\href@noop
  {} {\bibfield  {journal} {\bibinfo  {journal} {Phys. Rev. A}\ }\textbf
  {\bibinfo {volume} {80}},\ \bibinfo {pages} {021603} (\bibinfo {year}
  {2009})}\BibitemShut {NoStop}%
\bibitem [{\citenamefont {Pilati}\ and\ \citenamefont
  {Varma}(2017)}]{pilati2017localization}%
  \BibitemOpen
  \bibfield  {author} {\bibinfo {author} {\bibfnamefont {S.}~\bibnamefont
  {Pilati}}\ and\ \bibinfo {author} {\bibfnamefont {V.~K.}\ \bibnamefont
  {Varma}},\ }\href@noop {} {\bibfield  {journal} {\bibinfo  {journal} {Phys.
  Rev. A}\ }\textbf {\bibinfo {volume} {95}},\ \bibinfo {pages} {013613}
  (\bibinfo {year} {2017})}\BibitemShut {NoStop}%
\bibitem [{\citenamefont {Gordillo}\ \emph {et~al.}(2015)\citenamefont
  {Gordillo}, \citenamefont {Carbonell-Coronado},\ and\ \citenamefont
  {De~Soto}}]{PhysRevA.91.043618}%
  \BibitemOpen
  \bibfield  {author} {\bibinfo {author} {\bibfnamefont {M.~C.}\ \bibnamefont
  {Gordillo}}, \bibinfo {author} {\bibfnamefont {C.}~\bibnamefont
  {Carbonell-Coronado}}, \ and\ \bibinfo {author} {\bibfnamefont
  {F.}~\bibnamefont {De~Soto}},\ }\href@noop {} {\bibfield  {journal} {\bibinfo
   {journal} {Phys. Rev. A}\ }\textbf {\bibinfo {volume} {91}},\ \bibinfo
  {pages} {043618} (\bibinfo {year} {2015})}\BibitemShut {NoStop}%
\bibitem [{\citenamefont {Ma}\ \emph {et~al.}(2012)\citenamefont {Ma},
  \citenamefont {Pilati}, \citenamefont {Troyer},\ and\ \citenamefont
  {Dai}}]{dft}%
  \BibitemOpen
  \bibfield  {author} {\bibinfo {author} {\bibfnamefont {P.~N.}\ \bibnamefont
  {Ma}}, \bibinfo {author} {\bibfnamefont {S.}~\bibnamefont {Pilati}}, \bibinfo
  {author} {\bibfnamefont {M.}~\bibnamefont {Troyer}}, \ and\ \bibinfo {author}
  {\bibfnamefont {X.}~\bibnamefont {Dai}},\ }\href@noop {} {\bibfield
  {journal} {\bibinfo  {journal} {Nat. Phys.}\ }\textbf {\bibinfo {volume}
  {8}},\ \bibinfo {pages} {601} (\bibinfo {year} {2012})}\BibitemShut {NoStop}%
\bibitem [{\citenamefont {Pilati}\ \emph {et~al.}(2014)\citenamefont {Pilati},
  \citenamefont {Zintchenko},\ and\ \citenamefont {Troyer}}]{pilati2014}%
  \BibitemOpen
  \bibfield  {author} {\bibinfo {author} {\bibfnamefont {S.}~\bibnamefont
  {Pilati}}, \bibinfo {author} {\bibfnamefont {I.}~\bibnamefont {Zintchenko}},
  \ and\ \bibinfo {author} {\bibfnamefont {M.}~\bibnamefont {Troyer}},\
  }\href@noop {} {\bibfield  {journal} {\bibinfo  {journal} {Phys. Rev. Lett.}\
  }\textbf {\bibinfo {volume} {112}},\ \bibinfo {pages} {015301} (\bibinfo
  {year} {2014})}\BibitemShut {NoStop}%
\bibitem [{\citenamefont {Olshanii}(1998)}]{olshanii1998atomic}%
  \BibitemOpen
  \bibfield  {author} {\bibinfo {author} {\bibfnamefont {M.}~\bibnamefont
  {Olshanii}},\ }\href@noop {} {\bibfield  {journal} {\bibinfo  {journal}
  {Phys. Rev. Lett.}\ }\textbf {\bibinfo {volume} {81}},\ \bibinfo {pages}
  {938} (\bibinfo {year} {1998})}\BibitemShut {NoStop}%
\bibitem [{\citenamefont {Astrakharchik}\ \emph {et~al.}(2004)\citenamefont
  {Astrakharchik}, \citenamefont {Blume}, \citenamefont {Giorgini},\ and\
  \citenamefont {Pitaevskii}}]{astrakharchikgiorgini}%
  \BibitemOpen
  \bibfield  {author} {\bibinfo {author} {\bibfnamefont {G.}~\bibnamefont
  {Astrakharchik}}, \bibinfo {author} {\bibfnamefont {D.}~\bibnamefont
  {Blume}}, \bibinfo {author} {\bibfnamefont {S.}~\bibnamefont {Giorgini}}, \
  and\ \bibinfo {author} {\bibfnamefont {L.}~\bibnamefont {Pitaevskii}},\
  }\href@noop {} {\bibfield  {journal} {\bibinfo  {journal} {Phys. Rev. Lett.}\
  }\textbf {\bibinfo {volume} {93}},\ \bibinfo {pages} {050402} (\bibinfo
  {year} {2004})}\BibitemShut {NoStop}%
\bibitem [{tU()}]{tU}%
  \BibitemOpen
  \href@noop {} {}\bibinfo {note} {One has: $t = \int_0^{Ld}
  w_r^*(x)\left[\frac{\hbar^2}{2m} \frac{\partial^2}{\partial x^2} - v(x)
  \right] w_{r+1}(x) \mathrm{d} x $ and $U = g \int _0^{Ld}
  \left|w_r(x)\right|^4 \mathrm{d} x$, where $w_r(x)$ is the (lowest-band)
  Wannier orbital at site $r$.}\BibitemShut {Stop}%
\bibitem [{\citenamefont {Reynolds}\ \emph {et~al.}(1982)\citenamefont
  {Reynolds}, \citenamefont {Ceperley}, \citenamefont {Alder},\ and\
  \citenamefont {Lester~Jr}}]{reynolds1982fixed}%
  \BibitemOpen
  \bibfield  {author} {\bibinfo {author} {\bibfnamefont {P.~J.}\ \bibnamefont
  {Reynolds}}, \bibinfo {author} {\bibfnamefont {D.~M.}\ \bibnamefont
  {Ceperley}}, \bibinfo {author} {\bibfnamefont {B.~J.}\ \bibnamefont {Alder}},
  \ and\ \bibinfo {author} {\bibfnamefont {W.~A.}\ \bibnamefont {Lester~Jr}},\
  }\href@noop {} {\bibfield  {journal} {\bibinfo  {journal} {J. Chem. Phys.}\
  }\textbf {\bibinfo {volume} {77}},\ \bibinfo {pages} {5593} (\bibinfo {year}
  {1982})}\BibitemShut {NoStop}%
\bibitem [{\citenamefont {Ceperley}(1991)}]{ceperley1991fermion}%
  \BibitemOpen
  \bibfield  {author} {\bibinfo {author} {\bibfnamefont {D.~M.}\ \bibnamefont
  {Ceperley}},\ }\href@noop {} {\bibfield  {journal} {\bibinfo  {journal} {J.
  Stat. Phys.}\ }\textbf {\bibinfo {volume} {63}},\ \bibinfo {pages} {1237}
  (\bibinfo {year} {1991})}\BibitemShut {NoStop}%
\bibitem [{\citenamefont {Casula}\ \emph {et~al.}(2006)\citenamefont {Casula},
  \citenamefont {Sorella},\ and\ \citenamefont {Senatore}}]{casula2006ground}%
  \BibitemOpen
  \bibfield  {author} {\bibinfo {author} {\bibfnamefont {M.}~\bibnamefont
  {Casula}}, \bibinfo {author} {\bibfnamefont {S.}~\bibnamefont {Sorella}}, \
  and\ \bibinfo {author} {\bibfnamefont {G.}~\bibnamefont {Senatore}},\
  }\href@noop {} {\bibfield  {journal} {\bibinfo  {journal} {Phys. Rev. B}\
  }\textbf {\bibinfo {volume} {74}},\ \bibinfo {pages} {245427} (\bibinfo
  {year} {2006})}\BibitemShut {NoStop}%
\bibitem [{\citenamefont {Casula}\ \emph {et~al.}(2008)\citenamefont {Casula},
  \citenamefont {Ceperley},\ and\ \citenamefont {Mueller}}]{casula2008quantum}%
  \BibitemOpen
  \bibfield  {author} {\bibinfo {author} {\bibfnamefont {M.}~\bibnamefont
  {Casula}}, \bibinfo {author} {\bibfnamefont {D.}~\bibnamefont {Ceperley}}, \
  and\ \bibinfo {author} {\bibfnamefont {E.~J.}\ \bibnamefont {Mueller}},\
  }\href@noop {} {\bibfield  {journal} {\bibinfo  {journal} {Phys. Rev. A}\
  }\textbf {\bibinfo {volume} {78}},\ \bibinfo {pages} {033607} (\bibinfo
  {year} {2008})}\BibitemShut {NoStop}%
\bibitem [{\citenamefont {Matveeva}\ and\ \citenamefont
  {Astrakharchik}(2016{\natexlab{a}})}]{matveeva2016one}%
  \BibitemOpen
  \bibfield  {author} {\bibinfo {author} {\bibfnamefont {N.}~\bibnamefont
  {Matveeva}}\ and\ \bibinfo {author} {\bibfnamefont {G.}~\bibnamefont
  {Astrakharchik}},\ }\href@noop {} {\bibfield  {journal} {\bibinfo  {journal}
  {New J. Phys.}\ }\textbf {\bibinfo {volume} {18}},\ \bibinfo {pages} {065009}
  (\bibinfo {year} {2016}{\natexlab{a}})}\BibitemShut {NoStop}%
\bibitem [{\citenamefont {Boronat}(2002)}]{boronat}%
  \BibitemOpen
  \bibfield  {author} {\bibinfo {author} {\bibfnamefont {J.}~\bibnamefont
  {Boronat}},\ }in\ \href@noop {} {\emph {\bibinfo {booktitle} {Microscopic
  Approaches to Quantum Liquids in Confined Geometries}}},\ \bibinfo {editor}
  {edited by\ \bibinfo {editor} {\bibfnamefont {E.}~\bibnamefont {Krotscheck}}\
  and\ \bibinfo {editor} {\bibfnamefont {J.}~\bibnamefont {Navarro}}}\
  (\bibinfo  {publisher} {World Scientific, Singapore, 2002},\ \bibinfo {year}
  {2002})\ Chap.~\bibinfo {chapter} {2}, pp.\ \bibinfo {pages}
  {21--90}\BibitemShut {NoStop}%
\bibitem [{\citenamefont {Jaksch}\ \emph {et~al.}(1998)\citenamefont {Jaksch},
  \citenamefont {Bruder}, \citenamefont {Cirac}, \citenamefont {Gardiner},\
  and\ \citenamefont {Zoller}}]{jaksch}%
  \BibitemOpen
  \bibfield  {author} {\bibinfo {author} {\bibfnamefont {D.}~\bibnamefont
  {Jaksch}}, \bibinfo {author} {\bibfnamefont {C.}~\bibnamefont {Bruder}},
  \bibinfo {author} {\bibfnamefont {J.~I.}\ \bibnamefont {Cirac}}, \bibinfo
  {author} {\bibfnamefont {C.~W.}\ \bibnamefont {Gardiner}}, \ and\ \bibinfo
  {author} {\bibfnamefont {P.}~\bibnamefont {Zoller}},\ }\href@noop {}
  {\bibfield  {journal} {\bibinfo  {journal} {Phys. Rev. Lett.}\ }\textbf
  {\bibinfo {volume} {81}},\ \bibinfo {pages} {3108} (\bibinfo {year}
  {1998})}\BibitemShut {NoStop}%
\bibitem [{LWH()}]{LWHalfFilling}%
  \BibitemOpen
  \href@noop {} {}\bibinfo {note} {At half filling $N/L=1$, the Lieb-Wu result
  reads: $ E/N = -4t\int_0^{\infty} \frac{ J_0(x)J_1(x)}{x\left[1+\exp\left(x
  U/2t\right)\right]} \mathrm{d} x, $ where $J_i(x)$ are Bessel functions of
  first kind.}\BibitemShut {Stop}%
\bibitem [{\citenamefont {Xianlong}\ \emph {et~al.}(2006)\citenamefont
  {Xianlong}, \citenamefont {Polini}, \citenamefont {Asgari},\ and\
  \citenamefont {Tosi}}]{xianlong2006density}%
  \BibitemOpen
  \bibfield  {author} {\bibinfo {author} {\bibfnamefont {G.}~\bibnamefont
  {Xianlong}}, \bibinfo {author} {\bibfnamefont {M.}~\bibnamefont {Polini}},
  \bibinfo {author} {\bibfnamefont {R.}~\bibnamefont {Asgari}}, \ and\ \bibinfo
  {author} {\bibfnamefont {M.}~\bibnamefont {Tosi}},\ }\href@noop {} {\bibfield
   {journal} {\bibinfo  {journal} {Phys. Rev. A}\ }\textbf {\bibinfo {volume}
  {73}},\ \bibinfo {pages} {033609} (\bibinfo {year} {2006})}\BibitemShut
  {NoStop}%
\bibitem [{E2()}]{E2}%
  \BibitemOpen
  \href@noop {} {}\bibinfo {note} {One has: $E_2/(2\hbar \omega) =
  1/4+\epsilon/2$, where $\omega=2\sqrt{E_r V}/\hbar$ and $\epsilon \in [1/2,
  3/2]$ is the root of the transcendental equation:
  $\frac{\Gamma\left(3/4-\epsilon/2 \right)}{\Gamma\left(1/4-\epsilon/2
  \right)} = -\frac{\lambda}{2\sqrt{2}},$ where $\Gamma(x)$ is the Euler Gamma
  function and $\lambda =-2\sqrt{\hbar/m\omega}/a_{1D}$.}\BibitemShut {Stop}%
\bibitem [{\citenamefont {Krutitsky}(2016)}]{krutitsky2016ultracold}%
  \BibitemOpen
  \bibfield  {author} {\bibinfo {author} {\bibfnamefont {K.~V.}\ \bibnamefont
  {Krutitsky}},\ }\href@noop {} {\bibfield  {journal} {\bibinfo  {journal}
  {Physics Reports}\ }\textbf {\bibinfo {volume} {607}},\ \bibinfo {pages} {1}
  (\bibinfo {year} {2016})}\BibitemShut {NoStop}%
\bibitem [{\citenamefont
  {M{\"u}ller-Kirsten}(2006)}]{ma14ller2006introduction}%
  \BibitemOpen
  \bibfield  {author} {\bibinfo {author} {\bibfnamefont {H.~J.}\ \bibnamefont
  {M{\"u}ller-Kirsten}},\ }\href@noop {} {\emph {\bibinfo {title} {Introduction
  to Quantum Mechanics: Schr{\"o}dinger Equation and Path Integral}}}\
  (\bibinfo  {publisher} {World Scientific Publishing Co Inc},\ \bibinfo {year}
  {2006})\BibitemShut {NoStop}%
\bibitem [{\citenamefont {White}(1992)}]{white1992density}%
  \BibitemOpen
  \bibfield  {author} {\bibinfo {author} {\bibfnamefont {S.~R.}\ \bibnamefont
  {White}},\ }\href@noop {} {\bibfield  {journal} {\bibinfo  {journal} {Phys.
  Rev. Lett.}\ }\textbf {\bibinfo {volume} {69}},\ \bibinfo {pages} {2863}
  (\bibinfo {year} {1992})}\BibitemShut {NoStop}%
\bibitem [{DMR()}]{DMRG}%
  \BibitemOpen
  \href@noop {} {}\bibinfo {note} {The DMRG simulations are performed by using
  up to 1024 DMRG states and 5 finite size sweeps.}\BibitemShut {Stop}%
\bibitem [{\citenamefont {Schulz}(1990)}]{schulz1990correlation}%
  \BibitemOpen
  \bibfield  {author} {\bibinfo {author} {\bibfnamefont {H.}~\bibnamefont
  {Schulz}},\ }\href@noop {} {\bibfield  {journal} {\bibinfo  {journal} {Phys.
  Rev. Lett.}\ }\textbf {\bibinfo {volume} {64}},\ \bibinfo {pages} {2831}
  (\bibinfo {year} {1990})}\BibitemShut {NoStop}%
\bibitem [{\citenamefont {Frahm}\ and\ \citenamefont
  {Korepin}(1990)}]{frahm1990critical}%
  \BibitemOpen
  \bibfield  {author} {\bibinfo {author} {\bibfnamefont {H.}~\bibnamefont
  {Frahm}}\ and\ \bibinfo {author} {\bibfnamefont {V.}~\bibnamefont
  {Korepin}},\ }\href@noop {} {\bibfield  {journal} {\bibinfo  {journal} {Phy.
  Rev. B}\ }\textbf {\bibinfo {volume} {42}},\ \bibinfo {pages} {10553}
  (\bibinfo {year} {1990})}\BibitemShut {NoStop}%
\bibitem [{\citenamefont {Parola}\ and\ \citenamefont
  {Sorella}(1990)}]{parola1990asymptotic}%
  \BibitemOpen
  \bibfield  {author} {\bibinfo {author} {\bibfnamefont {A.}~\bibnamefont
  {Parola}}\ and\ \bibinfo {author} {\bibfnamefont {S.}~\bibnamefont
  {Sorella}},\ }\href@noop {} {\bibfield  {journal} {\bibinfo  {journal} {Phys.
  Rev. Lett.}\ }\textbf {\bibinfo {volume} {64}},\ \bibinfo {pages} {1831}
  (\bibinfo {year} {1990})}\BibitemShut {NoStop}%
\bibitem [{\citenamefont {Imada}\ \emph {et~al.}(1992)\citenamefont {Imada},
  \citenamefont {Furukawa},\ and\ \citenamefont
  {M.~Rice}}]{imada1992crossover}%
  \BibitemOpen
  \bibfield  {author} {\bibinfo {author} {\bibfnamefont {M.}~\bibnamefont
  {Imada}}, \bibinfo {author} {\bibfnamefont {N.}~\bibnamefont {Furukawa}}, \
  and\ \bibinfo {author} {\bibfnamefont {T.}~\bibnamefont {M.~Rice}},\
  }\href@noop {} {\bibfield  {journal} {\bibinfo  {journal} {The Physical
  Society of Japan}\ }\textbf {\bibinfo {volume} {61}},\ \bibinfo {pages}
  {3861} (\bibinfo {year} {1992})}\BibitemShut {NoStop}%
\bibitem [{\citenamefont {Sandvik}\ \emph {et~al.}(1993)\citenamefont
  {Sandvik}, \citenamefont {Scalapino},\ and\ \citenamefont
  {Singh}}]{sandvik1993quantum}%
  \BibitemOpen
  \bibfield  {author} {\bibinfo {author} {\bibfnamefont {A.}~\bibnamefont
  {Sandvik}}, \bibinfo {author} {\bibfnamefont {D.}~\bibnamefont {Scalapino}},
  \ and\ \bibinfo {author} {\bibfnamefont {C.}~\bibnamefont {Singh}},\
  }\href@noop {} {\bibfield  {journal} {\bibinfo  {journal} {Phys. Rev. B}\
  }\textbf {\bibinfo {volume} {48}},\ \bibinfo {pages} {2112} (\bibinfo {year}
  {1993})}\BibitemShut {NoStop}%
\bibitem [{\citenamefont {Qin}\ \emph {et~al.}(1995)\citenamefont {Qin},
  \citenamefont {Liang}, \citenamefont {Su},\ and\ \citenamefont
  {Yu}}]{qin1995density}%
  \BibitemOpen
  \bibfield  {author} {\bibinfo {author} {\bibfnamefont {S.}~\bibnamefont
  {Qin}}, \bibinfo {author} {\bibfnamefont {S.}~\bibnamefont {Liang}}, \bibinfo
  {author} {\bibfnamefont {Z.}~\bibnamefont {Su}}, \ and\ \bibinfo {author}
  {\bibfnamefont {L.}~\bibnamefont {Yu}},\ }\href@noop {} {\bibfield  {journal}
  {\bibinfo  {journal} {Phys. Rev. B}\ }\textbf {\bibinfo {volume} {52}},\
  \bibinfo {pages} {R5475} (\bibinfo {year} {1995})}\BibitemShut {NoStop}%
\bibitem [{\citenamefont {Iino}\ and\ \citenamefont
  {Imada}(1995)}]{iino1995crossover}%
  \BibitemOpen
  \bibfield  {author} {\bibinfo {author} {\bibfnamefont {Y.}~\bibnamefont
  {Iino}}\ and\ \bibinfo {author} {\bibfnamefont {M.}~\bibnamefont {Imada}},\
  }\href@noop {} {\bibfield  {journal} {\bibinfo  {journal} {J. Phys. Soc.
  Jpn}\ }\textbf {\bibinfo {volume} {64}},\ \bibinfo {pages} {4392} (\bibinfo
  {year} {1995})}\BibitemShut {NoStop}%
\bibitem [{\citenamefont {Schulz}(1993)}]{schulz1993wigner}%
  \BibitemOpen
  \bibfield  {author} {\bibinfo {author} {\bibfnamefont {H.}~\bibnamefont
  {Schulz}},\ }\href@noop {} {\bibfield  {journal} {\bibinfo  {journal} {Phys.
  Rev. Lett.}\ }\textbf {\bibinfo {volume} {71}},\ \bibinfo {pages} {1864}
  (\bibinfo {year} {1993})}\BibitemShut {NoStop}%
\bibitem [{HMc()}]{HMcorr}%
  \BibitemOpen
  \href@noop {} {}\bibinfo {note} {In the Hubbard model, one has:
  $g(l\equiv|l_1-l_2|)=-\frac{1}{(\pi l)^2} + B\cos\left(2k_Fl\right)
  \frac{\ln^{1/2}\left(l\right)}{l^{1+\alpha}} + \dots$, where $k_F=\pi N/(2L)$
  and $B$ is a model dependent coefficient.}\BibitemShut {Stop}%
\bibitem [{\citenamefont {Volosniev}\ \emph {et~al.}(2014)\citenamefont
  {Volosniev}, \citenamefont {Fedorov}, \citenamefont {Jensen}, \citenamefont
  {Valiente},\ and\ \citenamefont {Zinner}}]{volosniev2014strongly}%
  \BibitemOpen
  \bibfield  {author} {\bibinfo {author} {\bibfnamefont {A.}~\bibnamefont
  {Volosniev}}, \bibinfo {author} {\bibfnamefont {D.~V.}\ \bibnamefont
  {Fedorov}}, \bibinfo {author} {\bibfnamefont {A.~S.}\ \bibnamefont {Jensen}},
  \bibinfo {author} {\bibfnamefont {M.}~\bibnamefont {Valiente}}, \ and\
  \bibinfo {author} {\bibfnamefont {N.~T.}\ \bibnamefont {Zinner}},\
  }\href@noop {} {\bibfield  {journal} {\bibinfo  {journal} {Nat. Commun.}\
  }\textbf {\bibinfo {volume} {5}},\ \bibinfo {pages} {5300} (\bibinfo {year}
  {2014})}\BibitemShut {NoStop}%
\bibitem [{\citenamefont {Matveeva}\ and\ \citenamefont
  {Astrakharchik}(2016{\natexlab{b}})}]{Matveeva}%
  \BibitemOpen
  \bibfield  {author} {\bibinfo {author} {\bibfnamefont {N.}~\bibnamefont
  {Matveeva}}\ and\ \bibinfo {author} {\bibfnamefont {G.~E.}\ \bibnamefont
  {Astrakharchik}},\ }\href@noop {} {\bibfield  {journal} {\bibinfo  {journal}
  {New J. Phys.}\ }\textbf {\bibinfo {volume} {18}},\ \bibinfo {pages} {065009}
  (\bibinfo {year} {2016}{\natexlab{b}})}\BibitemShut {NoStop}%
\bibitem [{\citenamefont {Lindgren}\ \emph {et~al.}(2014)\citenamefont
  {Lindgren}, \citenamefont {Rotureau}, \citenamefont {Forss{\'e}n},
  \citenamefont {Volosniev},\ and\ \citenamefont {Zinner}}]{Lindgren}%
  \BibitemOpen
  \bibfield  {author} {\bibinfo {author} {\bibfnamefont {E.~J.}\ \bibnamefont
  {Lindgren}}, \bibinfo {author} {\bibfnamefont {J.}~\bibnamefont {Rotureau}},
  \bibinfo {author} {\bibfnamefont {C.}~\bibnamefont {Forss{\'e}n}}, \bibinfo
  {author} {\bibfnamefont {A.~G.}\ \bibnamefont {Volosniev}}, \ and\ \bibinfo
  {author} {\bibfnamefont {N.~T.}\ \bibnamefont {Zinner}},\ }\href@noop {}
  {\bibfield  {journal} {\bibinfo  {journal} {New J. Phys.}\ }\textbf {\bibinfo
  {volume} {16}},\ \bibinfo {pages} {063003} (\bibinfo {year}
  {2014})}\BibitemShut {NoStop}%
\bibitem [{\citenamefont {Bugnion}\ and\ \citenamefont
  {Conduit}(2013)}]{PhysRevA.87.060502}%
  \BibitemOpen
  \bibfield  {author} {\bibinfo {author} {\bibfnamefont {P.~O.}\ \bibnamefont
  {Bugnion}}\ and\ \bibinfo {author} {\bibfnamefont {G.~J.}\ \bibnamefont
  {Conduit}},\ }\href {\doibase 10.1103/PhysRevA.87.060502} {\bibfield
  {journal} {\bibinfo  {journal} {Phys. Rev. A}\ }\textbf {\bibinfo {volume}
  {87}},\ \bibinfo {pages} {060502} (\bibinfo {year} {2013})}\BibitemShut
  {NoStop}%
\bibitem [{\citenamefont {Gharashi}\ and\ \citenamefont
  {Blume}(2013)}]{PhysRevLett.111.045302}%
  \BibitemOpen
  \bibfield  {author} {\bibinfo {author} {\bibfnamefont {S.~E.}\ \bibnamefont
  {Gharashi}}\ and\ \bibinfo {author} {\bibfnamefont {D.}~\bibnamefont
  {Blume}},\ }\href {\doibase 10.1103/PhysRevLett.111.045302} {\bibfield
  {journal} {\bibinfo  {journal} {Phys. Rev. Lett.}\ }\textbf {\bibinfo
  {volume} {111}},\ \bibinfo {pages} {045302} (\bibinfo {year}
  {2013})}\BibitemShut {NoStop}%
\bibitem [{\citenamefont {Yang}\ \emph {et~al.}(2015)\citenamefont {Yang},
  \citenamefont {Guan},\ and\ \citenamefont {Pu}}]{PhysRevA.91.043634}%
  \BibitemOpen
  \bibfield  {author} {\bibinfo {author} {\bibfnamefont {L.}~\bibnamefont
  {Yang}}, \bibinfo {author} {\bibfnamefont {L.}~\bibnamefont {Guan}}, \ and\
  \bibinfo {author} {\bibfnamefont {H.}~\bibnamefont {Pu}},\ }\href {\doibase
  10.1103/PhysRevA.91.043634} {\bibfield  {journal} {\bibinfo  {journal} {Phys.
  Rev. A}\ }\textbf {\bibinfo {volume} {91}},\ \bibinfo {pages} {043634}
  (\bibinfo {year} {2015})}\BibitemShut {NoStop}%
\bibitem [{\citenamefont {Gharashi}\ \emph {et~al.}(2014)\citenamefont
  {Gharashi}, \citenamefont {Yin},\ and\ \citenamefont
  {Blume}}]{PhysRevA.89.023603}%
  \BibitemOpen
  \bibfield  {author} {\bibinfo {author} {\bibfnamefont {S.~E.}\ \bibnamefont
  {Gharashi}}, \bibinfo {author} {\bibfnamefont {X.~Y.}\ \bibnamefont {Yin}}, \
  and\ \bibinfo {author} {\bibfnamefont {D.}~\bibnamefont {Blume}},\ }\href
  {\doibase 10.1103/PhysRevA.89.023603} {\bibfield  {journal} {\bibinfo
  {journal} {Phys. Rev. A}\ }\textbf {\bibinfo {volume} {89}},\ \bibinfo
  {pages} {023603} (\bibinfo {year} {2014})}\BibitemShut {NoStop}%
\bibitem [{\citenamefont {Loft}\ \emph {et~al.}(2015)\citenamefont {Loft},
  \citenamefont {Dehkharghani}, \citenamefont {Mehta}, \citenamefont
  {Volosniev},\ and\ \citenamefont {Zinner}}]{loft2015variational}%
  \BibitemOpen
  \bibfield  {author} {\bibinfo {author} {\bibfnamefont {N.~J.~S.}\
  \bibnamefont {Loft}}, \bibinfo {author} {\bibfnamefont {A.~S.}\ \bibnamefont
  {Dehkharghani}}, \bibinfo {author} {\bibfnamefont {N.~P.}\ \bibnamefont
  {Mehta}}, \bibinfo {author} {\bibfnamefont {A.~G.}\ \bibnamefont
  {Volosniev}}, \ and\ \bibinfo {author} {\bibfnamefont {N.~T.}\ \bibnamefont
  {Zinner}},\ }\href@noop {} {\bibfield  {journal} {\bibinfo  {journal} {The
  European Physical Journal D}\ }\textbf {\bibinfo {volume} {69}},\ \bibinfo
  {pages} {65} (\bibinfo {year} {2015})}\BibitemShut {NoStop}%
\bibitem [{\citenamefont {Grining}\ \emph {et~al.}(2015)\citenamefont
  {Grining}, \citenamefont {Tomza}, \citenamefont {Lesiuk}, \citenamefont
  {Przybytek}, \citenamefont {Musia{\l}}, \citenamefont {Moszynski},
  \citenamefont {Lewenstein},\ and\ \citenamefont
  {Massignan}}]{grining2015crossover}%
  \BibitemOpen
  \bibfield  {author} {\bibinfo {author} {\bibfnamefont {T.}~\bibnamefont
  {Grining}}, \bibinfo {author} {\bibfnamefont {M.}~\bibnamefont {Tomza}},
  \bibinfo {author} {\bibfnamefont {M.}~\bibnamefont {Lesiuk}}, \bibinfo
  {author} {\bibfnamefont {M.}~\bibnamefont {Przybytek}}, \bibinfo {author}
  {\bibfnamefont {M.}~\bibnamefont {Musia{\l}}}, \bibinfo {author}
  {\bibfnamefont {R.}~\bibnamefont {Moszynski}}, \bibinfo {author}
  {\bibfnamefont {M.}~\bibnamefont {Lewenstein}}, \ and\ \bibinfo {author}
  {\bibfnamefont {P.}~\bibnamefont {Massignan}},\ }\href@noop {} {\bibfield
  {journal} {\bibinfo  {journal} {Phys. Rev. A}\ }\textbf {\bibinfo {volume}
  {92}},\ \bibinfo {pages} {061601} (\bibinfo {year} {2015})}\BibitemShut
  {NoStop}%
\bibitem [{\citenamefont {P\c{e}cak}\ \emph {et~al.}(2016)\citenamefont
  {P\c{e}cak}, \citenamefont {Gajda},\ and\ \citenamefont
  {Sowi{\'n}ski}}]{twoflavour}%
  \BibitemOpen
  \bibfield  {author} {\bibinfo {author} {\bibfnamefont {D.}~\bibnamefont
  {P\c{e}cak}}, \bibinfo {author} {\bibfnamefont {M.}~\bibnamefont {Gajda}}, \
  and\ \bibinfo {author} {\bibfnamefont {T.}~\bibnamefont {Sowi{\'n}ski}},\
  }\href@noop {} {\bibfield  {journal} {\bibinfo  {journal} {New J. Phys.}\
  }\textbf {\bibinfo {volume} {18}},\ \bibinfo {pages} {013030} (\bibinfo
  {year} {2016})}\BibitemShut {NoStop}%
\bibitem [{\citenamefont {Berger}\ \emph {et~al.}(2015)\citenamefont {Berger},
  \citenamefont {Anderson},\ and\ \citenamefont {Drut}}]{PhysRevA.91.053618}%
  \BibitemOpen
  \bibfield  {author} {\bibinfo {author} {\bibfnamefont {C.~E.}\ \bibnamefont
  {Berger}}, \bibinfo {author} {\bibfnamefont {E.~R.}\ \bibnamefont
  {Anderson}}, \ and\ \bibinfo {author} {\bibfnamefont {J.~E.}\ \bibnamefont
  {Drut}},\ }\href {\doibase 10.1103/PhysRevA.91.053618} {\bibfield  {journal}
  {\bibinfo  {journal} {Phys. Rev. A}\ }\textbf {\bibinfo {volume} {91}},\
  \bibinfo {pages} {053618} (\bibinfo {year} {2015})}\BibitemShut {NoStop}%
\bibitem [{\citenamefont {Dehkharghani}\ \emph {et~al.}(2016)\citenamefont
  {Dehkharghani}, \citenamefont {Volosniev},\ and\ \citenamefont
  {Zinner}}]{dehkharghani2016impenetrable}%
  \BibitemOpen
  \bibfield  {author} {\bibinfo {author} {\bibfnamefont {A.~S.}\ \bibnamefont
  {Dehkharghani}}, \bibinfo {author} {\bibfnamefont {A.~G.}\ \bibnamefont
  {Volosniev}}, \ and\ \bibinfo {author} {\bibfnamefont {N.~T.}\ \bibnamefont
  {Zinner}},\ }\href@noop {} {\bibfield  {journal} {\bibinfo  {journal}
  {Journal of Physics B: Atomic, Molecular and Optical Physics}\ }\textbf
  {\bibinfo {volume} {49}},\ \bibinfo {pages} {085301} (\bibinfo {year}
  {2016})}\BibitemShut {NoStop}%
\bibitem [{\citenamefont {Gharashi}\ \emph {et~al.}(2015)\citenamefont
  {Gharashi}, \citenamefont {Yin}, \citenamefont {Yan},\ and\ \citenamefont
  {Blume}}]{PhysRevA.91.013620}%
  \BibitemOpen
  \bibfield  {author} {\bibinfo {author} {\bibfnamefont {S.~E.}\ \bibnamefont
  {Gharashi}}, \bibinfo {author} {\bibfnamefont {X.~Y.}\ \bibnamefont {Yin}},
  \bibinfo {author} {\bibfnamefont {Y.}~\bibnamefont {Yan}}, \ and\ \bibinfo
  {author} {\bibfnamefont {D.}~\bibnamefont {Blume}},\ }\href {\doibase
  10.1103/PhysRevA.91.013620} {\bibfield  {journal} {\bibinfo  {journal} {Phys.
  Rev. A}\ }\textbf {\bibinfo {volume} {91}},\ \bibinfo {pages} {013620}
  (\bibinfo {year} {2015})}\BibitemShut {NoStop}%
\bibitem [{\citenamefont {Levinsen}\ \emph {et~al.}(2015)\citenamefont
  {Levinsen}, \citenamefont {Massignan}, \citenamefont {Bruun},\ and\
  \citenamefont {Parish}}]{Levinsene1500197}%
  \BibitemOpen
  \bibfield  {author} {\bibinfo {author} {\bibfnamefont {J.}~\bibnamefont
  {Levinsen}}, \bibinfo {author} {\bibfnamefont {P.}~\bibnamefont {Massignan}},
  \bibinfo {author} {\bibfnamefont {G.~M.}\ \bibnamefont {Bruun}}, \ and\
  \bibinfo {author} {\bibfnamefont {M.~M.}\ \bibnamefont {Parish}},\
  }\href@noop {} {\bibfield  {journal} {\bibinfo  {journal} {Science Advances}\
  }\textbf {\bibinfo {volume} {1}} (\bibinfo {year} {2015})}\BibitemShut
  {NoStop}%
\bibitem [{\citenamefont {Sowi\ifmmode~\acute{n}\else \'{n}\fi{}ski}\ \emph
  {et~al.}(2013)\citenamefont {Sowi\ifmmode~\acute{n}\else \'{n}\fi{}ski},
  \citenamefont {Grass}, \citenamefont {Dutta},\ and\ \citenamefont
  {Lewenstein}}]{PhysRevA.88.033607}%
  \BibitemOpen
  \bibfield  {author} {\bibinfo {author} {\bibfnamefont {T.}~\bibnamefont
  {Sowi\ifmmode~\acute{n}\else \'{n}\fi{}ski}}, \bibinfo {author}
  {\bibfnamefont {T.}~\bibnamefont {Grass}}, \bibinfo {author} {\bibfnamefont
  {O.}~\bibnamefont {Dutta}}, \ and\ \bibinfo {author} {\bibfnamefont
  {M.}~\bibnamefont {Lewenstein}},\ }\href@noop {} {\bibfield  {journal}
  {\bibinfo  {journal} {Phys. Rev. A}\ }\textbf {\bibinfo {volume} {88}},\
  \bibinfo {pages} {033607} (\bibinfo {year} {2013})}\BibitemShut {NoStop}%
\bibitem [{\citenamefont {Stella}\ \emph {et~al.}(2011)\citenamefont {Stella},
  \citenamefont {Attaccalite}, \citenamefont {Sorella},\ and\ \citenamefont
  {Rubio}}]{Stella}%
  \BibitemOpen
  \bibfield  {author} {\bibinfo {author} {\bibfnamefont {L.}~\bibnamefont
  {Stella}}, \bibinfo {author} {\bibfnamefont {C.}~\bibnamefont {Attaccalite}},
  \bibinfo {author} {\bibfnamefont {S.}~\bibnamefont {Sorella}}, \ and\
  \bibinfo {author} {\bibfnamefont {A.}~\bibnamefont {Rubio}},\ }\href@noop {}
  {\bibfield  {journal} {\bibinfo  {journal} {Phys. Rev. B}\ }\textbf {\bibinfo
  {volume} {84}},\ \bibinfo {pages} {245117} (\bibinfo {year}
  {2011})}\BibitemShut {NoStop}%
\end{thebibliography}
%
\end{document}